\documentclass{article}
\usepackage[numbers,sort&compress]{natbib}
\usepackage[utf8]{inputenc} 
\usepackage[T1]{fontenc}    

\usepackage{comment}
\usepackage{footnote}
\usepackage{graphicx}
\usepackage{amsmath}
\usepackage{caption}
\usepackage{subcaption}
\usepackage{listings}
\usepackage{amsmath}
\usepackage{hyperref}
\usepackage{multirow}
\usepackage{algorithm}
\usepackage[noend]{algpseudocode}
\usepackage{tikz}
\usetikzlibrary{positioning, arrows.meta, shapes.geometric, decorations.pathreplacing}

\usepackage{xcolor}

\newcommand{\theadb}[1]{\begin{tabular}{@{}c@{}}#1\end{tabular}}

\title{Celtibero: Robust Layered Aggregation for Federated Learning}
\author{Borja Molina Coronado\\\texttt{borja.molina@ehu.eus}}

\begin{document}
	
	\maketitle
	
	\begin{abstract}
		
		Federated Learning (FL) is an innovative approach to distributed machine learning. While FL offers significant privacy advantages, it also faces security challenges, particularly from poisoning attacks where adversaries deliberately manipulate local model updates to degrade model performance or introduce hidden backdoors. Existing defenses against these attacks have been shown to be effective when the data on the nodes is identically and independently distributed (i.i.d.), but they often fail under less restrictive, non-i.i.d data conditions. To overcome these limitations, we introduce Celtibero, a novel defense mechanism that integrates layered aggregation to enhance robustness against adversarial manipulation. Through extensive experiments on the MNIST and IMDB datasets, we demonstrate that Celtibero consistently achieves high main task accuracy (MTA) while maintaining minimal attack success rates (ASR) across a range of untargeted and targeted poisoning attacks. Our results highlight the superiority of Celtibero over existing defenses such as FL-Defender, LFighter, and FLAME, establishing it as a highly effective solution for securing federated learning systems against sophisticated poisoning attacks.

	\end{abstract}
	
	\section{Introduction}
	
	Federated Learning (FL) has emerged as a new machine learning paradigm for efficient distributed learning with privacy concerns \cite{mcmahan2017communication}. Under this paradigm, a central server coordinates the generation of a global model from a set of locally trained model in the client nodes. FL aims to address the data island problem, preventing the direct disclosure of local raw data on the clients \cite{zhang2021survey}. As no private data gets exchanged for training the models, FL posses an advance for applications in areas such as cybersecurity, healthcare or finance \cite{agrawal2022federated,rieke2020future,long2020federated,belenguer2023gowfed}, where strong privacy guarantees are fundamental.
	
	However, the FL framework is exposed to a wide variety of attacks, which can led to the leakage of clients' data or to harm the predictive capacity and reliability of the global model \cite{qammar2022federated}. In this context, the autonomy granted to clients in contributing to the final model also opens the door for adversarial peers to carry out poisoning attacks. In these attacks, through compromising the client's integrity or the communication channels between clients and the server adversaries inject poisoned data or malicious model updates \cite{rodriguez2023survey}. The consequences range from diminishing the efficacy of the global model to covertly crafting a model in such a way that its outcome can be controlled by the attacker \cite{shejwalkar2021manipulating, wang2020attack}.
	
	To address these threats, several defense mechanisms have been proposed to the FL framework. Existing defenses against poisoning attacks in federated learning settings can be broadly categorized into Influence Reduction (IR) and Detection and Filtering (DF) methods \cite{rieger2024crowdguard}. The first type of defenses aim to compute the global model using robust statistical measures, so that the effect of poisoned updates in the final aggregated model is minimal. Examples of this group include Krum \cite{blanchard2017machine}, median \cite{yin2018byzantine}, and Differential Privacy \cite{naseri2020local}. In contrast, detection and filtering mechanisms analyze model updates sent by the nodes to identify and filter out those suspicious of being poisoned \cite{tian2022comprehensive}. These mechanisms primarily rely on anomaly detectors based on unsupervised machine learning (ML) algorithms to identify local models that deviate from majority \cite{nguyen2022flame, jebreel2024lfighter}.
	
	Most existing defenses often present accurate and resilient models against attacks. However, some are based on assumptions that break the fundamentals of FL, such as the access to clean data on the server \cite{cao2020fltrust}, or propose modifications of the standard FL framework that may incur in increased communication overheads \cite{andreina2021baffle,qayyum2022making}. Other proposals instead, are devised for specific attack scenarios and may fail under different configurations. For example, defense methods that focus primarily on specific subsets of model parameters, such as the weights of the last hidden layer \cite{rieger2022deepsight, jebreel2023fl, jebreel2024lfighter, yang2024roseagg}, may not identify poisoning attacks that extend beyond these isolated layers and affect the entire neural network. Influence reduction (IR) methods may struggle to mitigate attacks when an attacker gains control over a sufficient number of federation participants to manipulate the statistical properties used in aggregation. Defenses based on differential privacy require finding an adequate balance of noise to effectively mitigate attacks while maintaining model accuracy on legitimate data, which may be scenario dependent \cite{dwork2006calibrating,naseri2020local}. Furthermore, most anomaly detectors used in detection and filtering methods to identify deviating updates may prove ineffective in scenarios with non-independent and identically distributed (non-i.i.d) data, as the presence of numerous distributions can lead to a diverse number of anomalies \cite{nguyen2022flame}. 
	
	In light of these challenges, we introduce Celtibero, a novel robust aggregator that integrates both detection and filtering, as well as influence reduction mechanisms. Unlike existing literature, Celtibero is not constrained to a limited set of weights extracted from specific layers of local models to perform aggregation. Instead, the comprehensive mechanism of Celtibero performs partial aggregations at each layer, constructing the global model as the combination of all benign partial aggregations. This flexibility allows it to adapt to diverse data distributions across different layers. Additionally, our defense is designed for a typical FL framework where participant nodes are only able to access the global model update sent by the server, making no-prior assumptions about the aggregation process and how the data is distributed in the nodes, i.e., whether data in the nodes is i.i.d or non-i.i.d. In summary, the main contributions of this work are summarized as follows:
	
	\begin{itemize}
		\item We introduce Celtibero\footnote{Celtibero refers to the ancient tribes that populated the Iberian Peninsula. The Celtiberians are known for their warrior culture and resistance to Roman conquest. They famously fought in the Numantine War, with the siege of Numantia in 133 BC being a notable event where they resisted Rome for an extended period. We named our defense mechanism in honor of their resilience against continuous attacks.}, a robust aggregator for Federated Learning (FL) that defends against poisoning attacks, effective in both i.i.d and non-i.i.d scenarios.
		\item We propose a novel layered approach to identify, filter out and reduce the influence of poisoned updates within the layers of local models, enabling robust aggregation in FL.
		\item Our extensive evaluations demonstrate that Celtibero effectively mitigates both untargeted and targeted poisoning attacks, outperforming state-of-the-art defenses.
	\end{itemize}
	
	This paper is organized as follows. Section \ref{related} analyzes the literature related to robust aggregation in FL. Section~\ref{background} gives background on the security concerns of FL and formalizes the threat model being considered in this work. The next section, Section \ref{solution}, describes the defense methodology of Celtibero. Section~\ref{experiments} presents our experimental setup and evaluates the robustness and performance of Celtibero, and compares it with several state-of-the-art methods. Finally, the main conclusions and future research lines of our work are discussed in Section~\ref{conclusions}.

	\section{Related Work}\label{related}
	
	Prior work aiming to counter-act poisoning attacks in FL includes methods that leverage robust statistics to reduce the impact of malicious models (IR) and anomaly detection schemes that aim to filter out anomalous models (DF). 
	
	\subsection{Influence Reduction defenses}
	
	In the first group, \cite{yin2018byzantine} employs the median of model parameters for aggregation to reduce the influence of anomalous outlier models. Krum \cite{blanchard2017machine} works by selecting the local model update that has the smallest sum of distances to the closest subset of other updates. As a variation of Krum, Median-Krum \cite{colosimo2023median}, computes the global model as the median of the local models with minimal sum of Euclidean distances to other local models. In \cite{naseri2020local}, the authors propose a differential privacy mechanism that adds random noise to the weights of local models with the aim of nullifying poisoned model weights. BaFFLe \cite{andreina2021baffle} identifies poisoned model updates based on client feedback for class prediction on the aggregated model. While median and Krum methods assume i.i.d. data and fail in non-i.i.d. scenarios, differential privacy require setting an adequate level of noise that mitigates attack influence without compromising model accuracy, but this is not trivial. The main drawback of BaFFLe is that it overlooks the fact that backdoor attacks are designed to avoid impacting the performance of models on legitimate data, making it challenging for this defense to identify backdoored models.
	
	FoolsGold \cite{fung2020limitations} assigns trust scores to models for weight averaging in aggregation, based on the assumption that malicious updates exhibit characteristics not present in legitimate updates. CONTRA \cite{awan2021contra} uses the pairwise cosine similarity between updates to limit the contribution of similar updates in the aggregation process, considering them potentially malicious. Similarly, RoseAgg \cite{yang2024roseagg} performs partial aggregation of local model parameters with similar directions. The global model is then computed based on the contribution of each partially aggregated model to the principal direction of all local weights, thereby reducing the influence of local models that deviate from the principal direction of the majority. FL-Defender \cite{jebreel2023fl} employs the two principal components of the cosine similarity matrix, derived from the gradients of the last layer of local models, to weight updates during aggregation. The aim is to mitigate the impact of malicious updates by reducing the influence of those whose projected angles significantly deviate from the median of the projections. All these methods, however, fall short when a large or small number of poisoned models are present in non-i.i.d. settings, or when poisoned models are crafted to closely resemble benign models in terms of their angles as in some backdoor attacks \cite{bagdasaryan2020backdoor}.
	
	\subsection{Detection and Filtering defenses}
	
	Defenses on the second group aim to detect an remove suspicious model updates. Auror \cite{shen2016auror} arranges local models into two groups based on the distribution of their weights, using for aggregation the models lying on the biggest cluster. DeepSight \cite{rieger2022deepsight} groups local models based on their label distribution as represented by the last layer parameters of local models. The goal is to determine local models with highly imbalanced label distributions as poisoned. FLAME \cite{nguyen2022flame} acts in two ways, first an outlier detection scheme is used to filter out abnormal local models. After that, the remaining models are clipped and noised before aggregation. LFighter groups local models into two clusters and identifies those lying in the more dense and smallest cluster as poisoned \cite{jebreel2024lfighter}. The main drawback of these methods lies in the assumption that benign local models are trained with i.i.d data, presenting limited ability in detecting poisoning attacks under non-i.i.d scenarios.
	
	To cope with non-i.i.d data, in \cite{qayyum2022making} an additional model that learns the association between the local model parameters and the training data is generated on each client. Finally this model is used to identify poisoned updates that follow a specific data poisoning behavior. Nonetheless, the need of this supplementary model highly increases the computational and communication costs of FL. Instead, CrowdGuard \cite{rieger2024crowdguard} proposes a validation mechanism on clients that is used as feedback to discard poisoned models before aggregation. However, it also assumes that each node can access other local models in the federation. This not only modifies the standard FL scheme and increases network overheads but also, could lead to craft poisoned models that target their closest local model to bypass detection. 	
	
	\section{Preliminaries}\label{background} 
	
	This section briefly describes the FL framework and presents poisoning attacks against FL. Finally, we describe the threat model assumed throughout this work. 
	
	\subsection{Federated Learning}
	
	In Federated Learning, an aggregator server builds a global model $G$ from local models $W_i$ sent by $K$ nodes $i \in \{1,...,K\}$ participating on the learning task. For each iteration of the FL process $t \in \{1,...,T\}$, the most typical scenario is to average local models to compute the global model $G^t = \sum_i^n W_i^t/n $. After aggregation, the server shares the global model parameters $G^t$ with the local nodes, so that they incrementally train this model using their data to obtain $W^{t+1}$. This process is repeated for a number of iterations to reach convergence. 
	
	\subsection{Poisoning Attacks against FL}
	
	In the FL setting, a node or set of nodes controlled by an adversary, or leveraging data obtained from unreliable sources, can lead to poisoned models that compromise the global model. Poisoning attacks against federated learning (FL) can be categorized into untargeted and targeted based on their goal \cite{tian2022comprehensive}. 
	
	\paragraph{Untargeted poisoning attacks.}
	
	These attacks inject noisy data into the training process by randomly altering the labels of instances in the original dataset $D$ to create a poisoned version $D_{pois}$. As a consequence, the local model trained on $D_{pois}$ will exhibit poor performance for real data. Once, this model is introduced in the FL aggregation process, it perturbs model parameters learn on legitimate data with the aim to degrade the overall performance of the global model \cite{fang2020local}.
	
	\paragraph{Targeted poisoning attacks}
	
	Differently from untargeted attacks, targeted attacks seek to introduce biases or perturbations into the model to force misclassification of inputs towards a specific target class. In the label flipping attack, the label of all instances in dataset $D$ with source class $c_{s}$ are changed to the class target class $c_{t}$. The purpose is to achieve misclassification towards class $c_t$ on the model trained with $D_{pois}$ for instances pertaining to the class $c_{s}$ \cite{biggio2012poisoning,tolpegin2020data}. During the FL iterative process, the poisoned model is sent for aggregation in order to disrupt the predictions of the global model for class $c_{s}$.
	
	As another class of targeted poisoning attacks, backdoor attacks implant a trigger into the model to manipulate its output towards a specific target class. To achieve this, the adversary assigns a target label and injects a specific (backdoor) pattern into the features of instances in the training dataset. The goal is for the trained model to classify instances with the backdoor pattern as the adversary-chosen target class, while performing normally on instances without the backdoor pattern (legitimate). Therefore, since backdoor attacks are designed not to impact the performance on legitimate instances, they are particularly difficult to detect. In FL, Model Replacement \cite{bagdasaryan2020backdoor}, Distributed Backdoor \cite{xie2019dba}, and Neurotoxin attacks \cite{zhang2022neurotoxin} send poisoned model updates to the FL server in order to introduce a durable and stealth backdoor trigger in the aggregated global model. 
	
	\subsubsection{Threat model}
	
	As in many other studies in the area, this work is based on some assumptions about the scope of the attacker. In this study, the goal of the attacker is to manipulate the global model to control its output or cause a high misclassification rate that renders the model useless. This is achieved by sending arbitrarily poisoned model updates during the FL operation. To do this, we consider an attacker that can control at most \( K' \) nodes, with \( K' < \frac{K}{2} \), where \( K \) is the total number of participating nodes in the FL process. In the controlled nodes, the attacker is able to manipulate the training data, as well as the training processes and model parameters sent and received during the FL operation. Furthermore, we assume that the aggregator server and other client nodes are secure, meaning that the attacker has no knowledge of the aggregation mechanism used by the server, nor access to the data, parameters, and training processes in the remaining nodes of the federation.
	
	\section{Defense Methodology: Celtibero}\label{solution}
	
	We design our defense, named Celtibero, to counter the effects of poisoned model updates during the aggregation process of FL conducted on the server. The goals of this method are:
	
	\emph{Effectiveness}. To identify poisoned models and eliminate their effect in the global model, preventing the adversary from compromising the model and achieving its goals. 
	
	\emph{Performance}. In the i.i.d scenario all local nodes have examples of all the classes in a similar proportion. In contrast, the non-i.i.d scenario assumes that local data sets can present different classes and proportions of examples of each class. Our defense mechanism should keep performance on the main task independently of the FL scenario presented.
	
	\emph{Robustness}. The defense mechanism does not make any assumption to specific attack conditions and is effective against poisoning attacks independently of their goal. This means that the defense should mitigate both, untargeted and targeted poisoning attacks.
	
	Celtibero combines detection and filtering with influence reduction to perform robust aggregation. The combination of these two mechanisms enhances the robustness of Celtibero against poisoning attacks. The detection and filtering mechanism aims to eliminate poisoned model updates from the aggregation process. This enables robust aggregation since malicious updates typically differ from the majority (benign) updates.
	
	However, some specific attacks may exploit this detection method by crafting updates that resemble benign updates. This aspect makes detection and filtering mechanisms prone to evasion, especially in non-iid environments where benign updates vary from one another. Therefore, putting in risk the global model that is being trained. To address this issue, Celtibero incorporates an influence reduction mechanism that minimizes the impact of undetected poisoned updates. Unlike detection and filtering mechanisms, which average the weights of updates identified as benign, our defense introduces a simple yet robust influence reduction method based on the median. Also, this mechanism provides robustness against updates that slightly modify the weights of models to perform the attack.
	
	Another characteristic of Celtibero is that the two mechanisms operate at the layer level. Contrary to defense approaches that consider the entire set of weights of models, the layered operation of Celtibero is especially relevant for combating attacks that spread through various layers of a model to achieve stealthiness. By accurately identifying and removing a poisoned update at a specific layer, Celtibero can break the poisoning pattern, potentially neutralizing the impact of attacks even if the malicious pattern is not fully detected across all layers. Moreover, the layered approach allows for the isolation of layer-specific poisoning patterns, which might be overlooked when considering the entire weight set collectively. This approach enables a more precise and effective defense compared to mechanisms that only analyze weights from a specific layer or from all layers at once. The next section specifies how all these mechanisms are integrated into Celtibero.
	
	\subsection{Layered Detection, Filterng and Influence Reduction Process}
	
	\begin{algorithm}
		\caption{Celtibero Defense Procedure}\label{celtibero-alg}
		\begin{algorithmic}[1]
			\Require Global model weights $W_G$, Local model weights $W_L^i$ for each node $i$, Number of layers $L$
			\Ensure Aggregated global model weights $W_G'$
			
			\For{each layer $l$ in $1$ to $L$}
			\State $\Delta W_L^{i,l} = W_L^{i,l} - W_G^l$ 
			
			\State $D^{i,j} = \text{cosine\_distance}(\Delta W_L^{i,l}, \Delta W_L^{j,l})$
			
			\State \text{Perform agglomerative clustering on } $\{\Delta W_L^{i,l}\}$ \text{ using } $D^{i,j}$ \text{ to form two clusters } $C_1$ \text{ and } $C_2$
			
			\State $density(C_k) = \frac{2}{|C_k|(|C_k|-1)} \sum_{\substack{i,j \in C_k \\ i \neq j}} D^{i,j}$ \text{ for } $k = 1, 2$
			
			\State $score(C_k) = |C_k| \cdot density(C_k)$ \text{ for } $k = 1, 2$
			
			\If{$score(C_1) < score(C_2)$}
			\State \text{Label } $C_1$ \text{ as poisoned and } $C_2$ \text{ as benign}
			\Else
			\State \text{Label } $C_2$ \text{ as poisoned and } $C_1$ \text{ as benign}
			\EndIf
			
			\State $\Delta W_{benign}^{l} = \{\Delta W_L^{i,l} \mid \Delta W_L^{i,l} \in \text{benign cluster}\}$
			
			\State $W_G^{l'} = W_G^l + median(\Delta W_{benign}^{l})$
			\EndFor
			
			\State $W_G' = \{W_G^{1'}, W_G^{2'}, \ldots, W_G^{L'}\}$
			
			\State \Return $W_G'$
		\end{algorithmic}
	\end{algorithm}
	
	The detection procedure of Celtibero is detailed in Algorithm \ref{celtibero-alg}. First, Celtibero computes the weight updates or gradients ($\Delta W_L^{i,l}$) at each layer based on the difference between the weights of the global model at the previous iteration and the weights of local models at the current iteration (line 2). The gradient updates of models can be interpreted as vectors that alter the weights of the previous global model in two distinct ways: by altering their magnitude (verified by Euclidean distance) or by changing the direction of the vector (reflected in increased cosine distance) relative to the previous weight state. To characterize local models for clustering, Celtibero relies on the cosine pairwise distances between the gradient updates (line 3) as they are useful to exhibit update patterns of local models. 
	
	To discern between benign and poisoned gradient vectors at each layer based on their angle contribution as indicated by the distance matrix $D$, Celtibero uses agglomerative clustering \cite{hastie2009elements}. In this regard, we hypothesize that poisoned updates will always share a training objective, which is to poison the global model. In contrast, the training objective of legitimate models may differ depending of their label distribution. Therefore, we argue that a stronger relation will always be present among poisoned model updates independently of the distribution of the data used for training in the local nodes. This relation can be captured by the hierarchical process followed by agglomerative clustering, since it progressively merges the pairs of gradient vectors with the closest angle contribution until all of them are grouped into two clusters.
	
	Once gradient vectors have been grouped into two clusters, Celtibero scores the clusters according to their size and density to label them as either benign or poisoned. First, Celtibero computes the density of each cluster (line 5) based on the average pairwise cosine distances between every pair of gradient vectors within the cluster. Then, the cluster density is proportionally weighted based on the size of the clusters (line 6). As mentioned, since poisoned updates will share a common attack objective, they will tend to form smaller and denser clusters. Therefore, the cluster with the smaller score is identified as poisoned by Celtibero, whereas gradient updates lying in the cluster with higher score are considered benign (lines 7-10). 
	
	Finally, the influence reduction process of Celtibero aggregates the benign gradient updates by taking the median value of each weight (line 12). The result of this operation is a partial aggregation of benign weights $W_G^{l'}$ at layer $l$ of the model. By repeating the entire process for each of the individual layers of the model, Celtibero performs layered aggregation. Finally, the partial aggregations for all layers are combined to form the global model $W_G'$ update.

	\section{Experimentation}\label{experiments}
	
	This section presents the evaluation of Celtibero to show its effectiveness against poisoning attacks in comparison to six state-of-the-art defenses: Median-Krum \cite{colosimo2023median}, foolsgold \cite{fung2020limitations}, FL-Defender \cite{jebreel2023fl}, FLAME \cite{nguyen2022flame}, LFighter \cite{jebreel2024lfighter} and RoseAgg \cite{yang2024roseagg}. We conduct all experiments using the Tensorflow deep learning framework \cite{tensorflow2015-whitepaper}. Our code is publicly available at \url{gitlab-borja}\footnote{We will publish the link upon paper acceptance}.
	
	\subsection{Experimental Setup}
	
	\begin{table*}[t]
		\centering
		\caption{Datasets and models used in our evaluations}
		\resizebox{\textwidth}{!}{%
			\begin{tabular}{  l | c | c | c | c | c | c }
				Task & Dataset & \theadb{\#training\\samples} & \theadb{\#testing\\samples} & Model & \#params & \theadb{Target\\class} \\ \hline
				\multirow{1}{*}{Image classification} & MNIST & 60K & 10K & CNN & $\sim$165K & 0 \\ 
				Sentiment analysis & IMDB & 25K & 25K & BiLSTM & $\sim$350K & positive \\ \hline
			\end{tabular}
		}%
		\label{tab:DLmodels}
	\end{table*}
	
	\paragraph{Datastets} To assess the effectivenes of Celtibero during our evaluations, we consider two benchmark datasets typically used in the literature. Specifically, we use the MNIST \cite{lecun1998mnist} digit recognition dataset, which consist of 70k handwritten digit images from 0 to 9; and the IMDB Large Movie Review dataset \cite{maas2011learning}, formed by 50k movie reviews and their corresponding sentiment binary labels. The models used for each of these datasets are outlined in Table~\ref{tab:DLmodels}
	
	\paragraph{FL setup} To simulate the FL framework, we partition the training datasets into local datasets $D_k \in \{D_1,...,D_K\}$, where each $D_k$ corresponds to a node $Q_k \in \{Q_1,...,Q_K\}$. In the i.i.d scenario, all $D_k$ have the same size and exhibit a similar class distribution. However, for non-i.i.d scenarios, we adopt a common approach used in the FL literature \cite{zhu2021federated,yang2024roseagg}, which manipulate the degree of non-i.i.d data across clients by adjusting the $\alpha$ parameter of the Dirichlet distribution \cite{lin2016dirichlet}. In our non-i.i.d experiments, we set $\alpha = 0.5$. We set the number of client nodes participating in the federation $K$ to 100 and 20 for the i.i.d and non-i.i.d scenarios, respectively, of which 40\% are controlled by an adversary to send poisoned updates to the server. We perform 50 FL training (aggregation) rounds for each experiment, with 3 local training epochs per round. Following \cite{nguyen2022flame}, we simulate a realistic scenario where in each round the number of clients that is available is randomly selected between 60\% and 90\% of the total nodes. Therefore, the number of adversarial nodes participating in each round also varies randomly.
	
	\paragraph{Attacks setup} We evaluate Celtibero against five untargeted and targeted poisoning attacks: Untargeted Label Flipping Attack (uLFA) \cite{tolpegin2020data}, Targeted Label Flipping Attack (tLFA) \cite{bhagoji2019analyzing}, Model Replacement Attack (MRA) \cite{bagdasaryan2020backdoor}, Distributed Backdoor Attack (DBA) \cite{xie2019dba} and Neurotoxin \cite{zhang2022neurotoxin}. Label flipping attacks, such as uLFA and tLFA, involve training malicious models with mislabeled data to poison the global model and render it unusable. Attacks such as MRA, DBA and Neurotoxin focus on introducing a backdoor pattern into the global model, allowing the attacker to produce predictions favoring a class chosen by the attacker. For MRA, DBA and Neurotoxin attacks, a random backdoor pattern is selected. In our experiments with targeted attacks, we assigned the target class for each dataset as indicated in the right column of Table~\ref{tab:DLmodels}.
	
	\paragraph{Evaluation Metrics} We consider two typically used metrics in the FL attack literature to evaluate the effectiveness of Celtibero. \emph{MTA (Main Task Accuracy)} refers to the accuracy of the model on the task it is designed for. It accounts for the proportion of samples that are predicted correctly by the model from all total predictions made by the model. \emph{ASR (Attack Success Rate)} measures the effectiveness of the poisoning attack under evaluation. Since label flipping attacks seek to harm the detection capacity of the model, for the uLFA, the ASR measures the decay ratio in the MTA accuracy with respect to the no-attack model, whereas for the tLFA, the ASR accounts for the decay on the accuracy of the source class with respect to the corresponding value of the no-attack model. In the case of backdoor attacks, the ASR indicates the proportion of poisoned samples with the backdoor that are identified by the model as the class intended by the attacker.	
	
	\subsection{Experimental Results}
	
	In this section, we show the effectiveness of Celtibero in detecting and mitigating poisoning attacks under i.i.d and non-i.i.d FL scenarios.
	
	\subsubsection{i.i.d data.}
	
	Table~\ref{tab:mnist-iid} presents the results of our experiments on the MNIST dataset under the i.i.d. scenario. The first row details the performance of the baseline model using FedAvg, i.e., without any defense mechanism. The baseline model shows an 8\% accuracy drop under untargeted label flipping attacks (uLFA), while the targeted version of this attack (tLFA) only affects accuracy by 1\% on the targeted class. This relatively minor impact contrasts sharply with the significant vulnerability of the model to backdoor attacks, where both DBA and MRA achieve a 98\% success rate. However, the Neurotoxin attack has a reduced impact, with only 10\% of attacks succeeding, likely due to the dilution of malicious updates by legitimate ones during aggregation.
	
	Among the seven robust aggregation defenses evaluated, four exhibit vulnerabilities to at least one targeted poisoning attack. RoseAgg is the most susceptible, with attack success rates of 97.6\% and 98.5\% for uLFA and MRA, respectively. FoolsGold also shows a significant weakness against MRA, with a 98.6\% success rate for backdoor attacks. Additionally, FLAME and MedianKrum are notably affected by the Neurotoxin attack, with 93.2\% and 91.5\% of poisoned samples succeeding, respectively. In contrast, Celtibero, FL-Defender, and LFighter demonstrate robustness across all five evaluated attacks, with Celtibero emerging as the most effective defense in terms of average MTA and ASR values.

	\begin{table*}[tb]
		\centering
		\caption{MNIST results for the i.i.d scenario}
		\resizebox{\textwidth}{!}{%
			\begin{tabular}{  l | c  c | c  c | c  c | c  c | c  c | c  c |}
				\multirow{2}{*}{Defenses} & \multicolumn{2}{|c|}{No attack} & \multicolumn{2}{|c|}{uLFA \cite{fang2020local}} & \multicolumn{2}{|c|}{tLFA \cite{bhagoji2019analyzing}} & \multicolumn{2}{|c|}{MRA \cite{bagdasaryan2020backdoor}} & \multicolumn{2}{|c|}{DBA \cite{xie2019dba}} & \multicolumn{2}{|c|}{Neurotoxin \cite{zhang2022neurotoxin}} \\ \cline{2-13} 
				& MTA & ASR & MTA & ASR & MTA & ASR & MTA & ASR & MTA & ASR & MTA & ASR \\ \hline
				\textit{No Defense} & 0.973 & - & 0.889 & 0.086 & 0.965 & 0.017 & 0.969 & 0.986 & 0.969 & 0.987 & 0.955 & 0.102 \\ \hline 
				\textit{MedianKrum \cite{colosimo2023median}} & 0.969 & - & 0.97 & 0 & 0.966 & 0.001 & 0.966 & 0 & 0.967 & 0 & 0.615 & 0.915 \\
				\textit{FoolsGold \cite{fung2020limitations}} & 0.973 & - & 0.964 & 0.009 & 0.964 & 0 & 0.968 & 0.986 & 0.962 & 0 & 0.962 & 0 \\
				\textit{RoseAgg \cite{yang2024roseagg}} & 0.972 & - & 0.023 & 0.976 & 0.974 & 0 & 0.970 & 0.985 & 0.973 & 0 & 0.967 & 0  \\
				\textit{FL-Defender \cite{jebreel2023fl}} & 0.962 & - & 0.959 & 0.003 & 0.963 & 0 & 0.959 & 0 & 0.960 & 0 & 0.959 & 0 \\
				\textit{FLAME \cite{nguyen2022flame}} & 0.969 & - & 0.970 & 0 & 0.970 & 0 & 0.968 & 0 & 0.969 & 0 & 0.895 & 0.932 \\
				\textit{LFighter \cite{jebreel2024lfighter}} & 0.974 & - & 0.969 & 0.004 & 0.971 & 0 & 0.968 & 0 & 0.970 & 0 & 0.969 & 0 \\ \hline
				\textit{Celtibero} & 0.973 & - & 0.972 & 0.001 & 0.971 & 0.001 & 0.968 & 0 & 0.969 & 0 & 0.967 & 0 \\
				
			\end{tabular}
		}%
		\label{tab:mnist-iid}
	\end{table*}
	
	When evaluated on the IMDB dataset (see Table~\ref{tab:imdb-iid}), the results are largely consistent with those observed for the MNIST dataset, with the notable exception of label flipping attacks. Due to the binary classification nature of the IMDB task, these attacks significantly degrade final task accuracy (MTA), reducing it to approximately 0.5 for models trained with vulnerable aggregation mechanisms such as FedAvg (baseline), MedianKrum, FL-Defender, and RoseAgg. Additionally, FLAME and FoolsGold exhibit classifier instability when faced with the Neurotoxin attack. In contrast, robust aggregation mechanisms such as LFighter and Celtibero demonstrate resilience across all attack types, maintaining high and stable MTA values and low ASR. However, Celtibero shows a slight vulnerability to the tLFA attack, with 18.9\% ASR.
	
	\begin{table*}[tb!]
		\centering
		\caption{IMDB results for the i.i.d scenario}
		\resizebox{\textwidth}{!}{%
			\begin{tabular}{  l | c  c | c  c | c  c | c  c | c  c | c  c |}
				\multirow{2}{*}{Defenses} & \multicolumn{2}{|c|}{No attack} & \multicolumn{2}{|c|}{uLFA \cite{fang2020local}} & \multicolumn{2}{|c|}{tLFA \cite{bhagoji2019analyzing}} & \multicolumn{2}{|c|}{MRA \cite{bagdasaryan2020backdoor}} & \multicolumn{2}{|c|}{DBA \cite{xie2019dba}} & \multicolumn{2}{|c|}{Neurotoxin \cite{zhang2022neurotoxin}} \\ \cline{2-13} 
				& MTA & ASR & MTA & ASR & MTA & ASR & MTA & ASR & MTA & ASR & MTA & ASR \\ \hline
				\textit{No Defense} & 0.875 & - & 0.501 & 0.426 & 0.510 & 0.972 & 0.873 & 0.752 & 0.849 & 0.797 & 0.863 & 0 \\ \hline 
				\textit{MedianKrum \cite{colosimo2023median}} & 0.865 & - & 0.5 & 0.422 & 0.5 & 1 & 0.851 & 0.030 & 0.866 & 0 & 0.5 & 0 \\
				\textit{FoolsGold \cite{fung2020limitations}} & 0.793 & - & 0.772 & 0.026 & 0.773 & 0.053 & 0.702 & 0.135 & 0.501 & 0 & 0.565 & 0 \\
				\textit{RoseAgg \cite{yang2024roseagg}} & 0.5 & - & 0.549 & 0 & 0.684 & 0.611 & 0.854 & 0.720 & 0.782 & 0.618 & 0.5 & 0 \\
				\textit{FL-Defender \cite{jebreel2023fl}} & 0.5 & - & 0.666 & 0 & 0.512 & 0.975 & 0.871 & 0.071 & 0.576 & 0.010 & 0.818 & 0 \\
				\textit{FLAME \cite{nguyen2022flame}} & 0.874 & - & 0.873 & 0 & 0.873 & 0 & 0.867 & 0.393 & 0.867 & 0 & 0.5 & 0 \\
				\textit{LFighter \cite{jebreel2024lfighter}} & 0.877 & - & 0.875 & 0.002 & 0.872 & 0.005 & 0.870 & 0.004 & 0.868 & 0 & 0.875 & 0 \\ \hline
				\textit{Celtibero} & 0.875 & - & 0.837 & 0.043 & 0.821 & 0.189 & 0.853 & 0.002 & 0.800 & 0.007 & 0.849 & 0.001 \\
				
			\end{tabular}
		}%
		\label{tab:imdb-iid}
	\end{table*}
	
	\subsubsection{Non-i.i.d data.} 
	
	Table~\ref{tab:mnist-niid} presents the results of our experiments on the MNIST dataset under the non-i.i.d. scenario. The baseline models trained with FedAvg are vulnerable to all attack types, resulting in fully compromised models. Notably, employing robust aggregators does not necessarily confer resilience against attacks since most defenses fail to maintain low Attack Success Rates (ASR), particularly against MRA, DBA, and Neurotoxin attacks. Even the robust mechanisms identified in the i.i.d. scenario, such as LFighter and FL-Defender, are compromised by MRA and DBA attacks, with ASR values reaching 99\%. Similarly, another popular defenses like MedianKrum and FLAME produce models successfully backdoored by MRA and Neurotoxin attacks, as indicated by their high ASR values. The only defense that proves effective across all attack types in this non-i.i.d. setting is Celtibero, which consistently keeps ASR below 2\% on average while maintaining high Main Task Accuracy (MTA) values of 97\%. Remarkably, Celtibero is the only defense that demonstrates complete robustness against tLFA, MRA, and Neurotoxin attacks, achieving 0\% ASR in these cases.

	\begin{table*}[t!]
		\centering
		\caption{MNIST results for the non-i.i.d. scenario}
		\resizebox{\textwidth}{!}{%
			\begin{tabular}{  l | c  c | c  c | c  c | c  c | c  c | c  c |}
				\multirow{2}{*}{Defenses} & \multicolumn{2}{|c|}{No attack} & \multicolumn{2}{|c|}{uLFA \cite{fang2020local}} & \multicolumn{2}{|c|}{tLFA \cite{bhagoji2019analyzing}} & \multicolumn{2}{|c|}{MRA \cite{bagdasaryan2020backdoor}} & \multicolumn{2}{|c|}{DBA \cite{xie2019dba}} & \multicolumn{2}{|c|}{Neurotoxin \cite{zhang2022neurotoxin}} \\ \cline{2-13} 
				& MTA & ASR & MTA & ASR & MTA & ASR & MTA & ASR & MTA & ASR & MTA & ASR \\ \hline
				\textit{No Defense} & 0.979  & - & 0.769 & 0.214 & 0.973 & 0.013 & 0.978 & 0.991 & 0.978 & 0.992 & 0.976 & 0.836 \\ \hline 
				\textit{MedianKrum \cite{colosimo2023median}} & 0.958 & - & 0.960 & 0 & 0.961 & 0 & 0.953 & 0.989 & 0.962 & 0 & 0.951 & 0.480 \\
				\textit{FoolsGold \cite{fung2020limitations}} & 0.980 & - & 0.211 & 0.784 & 0.977 & 0.002 & 0.978 & 0.991 & 0.979 & 0.992 & 0.976 & 0 \\
				\textit{RoseAgg \cite{yang2024roseagg}} & 0.962 & - & 0.947 & 0.016 & 0.969 & 0.056 & 0.974 & 0.988 & 0.968 & 0.983 & 0.977 & 0.989 \\
				\textit{FL-Defender \cite{jebreel2023fl}} & 0.974 & - & 0.896 & 0.080 & 0.975 & 0 & 0.975 & 0.989 & 0.977 & 0.992 & 0.970 & 0.007 \\
				\textit{FLAME \cite{nguyen2022flame}} & 0.970 & - & 0.975 & 0 & 0.975 & 0.001 & 0.969 & 0.991 & 0.976 & 0 & 0.970 & 0.981 \\
				\textit{LFighter \cite{jebreel2024lfighter}} & 0.979 & - & 0.916 & 0.064 & 0.977 & 0 & 0.979 & 0.991 & 0.980 & 0.990 & 0.977 & 0 \\ \hline
				\textit{Celtibero} & 0.977 & - & 0.945 & 0.032 & 0.975 & 0 & 0.974 & 0 & 0.974 & 0.046 & 0.973 & 0 \\
				
			\end{tabular}
		}%
		\label{tab:mnist-niid}
	\end{table*}
	
	\begin{table*}[t!]
		\centering
		\caption{IMDB results for the non-i.i.d scenario}
		\resizebox{\textwidth}{!}{%
			\begin{tabular}{  l | c  c | c  c | c  c | c  c | c  c | c  c |}
				\multirow{2}{*}{Defenses} & \multicolumn{2}{|c|}{No attack} & \multicolumn{2}{|c|}{uLFA \cite{fang2020local}} & \multicolumn{2}{|c|}{tLFA \cite{bhagoji2019analyzing}} & \multicolumn{2}{|c|}{MRA \cite{bagdasaryan2020backdoor}} & \multicolumn{2}{|c|}{DBA \cite{xie2019dba}} & \multicolumn{2}{|c|}{Neurotoxin \cite{zhang2022neurotoxin}} \\ \cline{2-13} 
				& MTA & ASR & MTA & ASR & MTA & ASR & MTA & ASR & MTA & ASR & MTA & ASR \\ \hline
				\textit{No Defense} & 0.849 & - & 0.513 & 0.395 & 0.508 & 0.978 & 0.847 & 0 & 0.839 & 0.819 & 0.823 & 0 \\ \hline 
				\textit{MedianKrum \cite{colosimo2023median}} & 0.640 & - & 0.5 & 0.219 & 0.502 & 0.957 & 0.765 & 0 & 0 & 0.003 & 0.581 & 0 \\
				\textit{FoolsGold \cite{fung2020limitations}} & 0.543 & - & 0.760 & 0 & 0.746 & 0 & 0.734 & 0 & 0.766 & 0.056 & 0.733 & 0 \\
				\textit{RoseAgg \cite{yang2024roseagg}} & 0.762 & - & 0.785 & 0 & 0.776 & 0.341 & 0.708 & 0 & 0.801 & 0.067 & 0.5 & 0 \\
				\textit{FL-Defender \cite{jebreel2023fl}} & 0.5 & - & 0.5 & 0 & 0.5 & 0 & 0.5 & 0 & 0.5 & 0  & 0.5 & 0 \\
				\textit{FLAME \cite{nguyen2022flame}} & 0.820 & - & 0.5 & 0.390 & 0.5 & 1 & 0.821 & 0 & 0.818 & 0.723 & 0.5 & 0 \\
				\textit{LFighter \cite{jebreel2024lfighter}} & 0.824 & - & 0.793 & 0.036 & 0.5 & 1 & 0.848 & 0 & 0.835 & 0.790 & 0.844 & 0 \\ \hline
				\textit{Celtibero} & 0.850 & - & 0.704 & 0.170 & 0.839 & 0.030 & 0.842 & 0 & 0.826 & 0.645 & 0.851 & 0 \\
				
			\end{tabular}
		}%
		\label{tab:imdb-niid}
	\end{table*}
	
	Results for the IMDB database for the non-i.i.d scenario are depicted in Table~\ref{tab:imdb-niid}. As can be seen, most robust aggregators in this scenario lead to unstable models, with Main Task Accuracy (MTA) values dropping to 0.5. This significant model degradation is caused by the aggregation mechanism when filtering out legitimate model updates. Only two mechanisms, FoolsGold and Celtibero, demonstrated stable performance on legitimate data. When comparing the two, FoolsGold yielded lower MTA values of 71.4\% on average, comparing to the obtained by Celtibero models of 81.9\%. Nonetheless, Celtibero is vulnerable to DBA attack with 64.5\% of successful attempts with respect to the ASR of 5.6\% obtained by FoolsGold.

	\section{Conclusions}\label{conclusions}
	
	In conclusion, our experimental results confirm that while existing defenses are primarily effective in FL scenarios with i.i.d. data, they often fail to prevent poisoning attacks under non-i.i.d. conditions. In many cases, these approaches resulted in low stability models, rendering them ineffective. To address this challenge, we proposed Celtibero, a novel defense mechanism based on layered aggregation through the examination of the angle contribution of updates using agglomerative clustering. Across the different experiments conducted, Celtibero consistently achieved high main task accuracy with minimal attack success rates against a range of untargeted and targeted poisoning attacks in both i.i.d. and non-i.i.d. scenarios. These results highlight that Celtibero is not only a highly effective defense strategy, outperforming current state-of-the-art defenses such as FL-Defender, LFighter, and FLAME, but also significantly enhances the security and reliability of the global model without compromising its stability.
	
	\bibliography{references}
	
	\bibliographystyle{unsrtnat}

\end{document}